\documentclass[%
 reprint,
groupedaddress,
 amsmath,amssymb,
 aps,
prb,
]{revtex4-1}

\usepackage{graphicx}
\usepackage{dcolumn}
\usepackage{bm}
\usepackage{xcolor}

\begin{document}

\newcommand{\be}{\begin{equation}}
\newcommand{\ee}[1]{\label{#1}\end{equation}}
\newcommand{\bem}{\begin{eqnarray}}
\newcommand{\eem}[1]{\label{#1}\end{eqnarray}}
\newcommand{\eq}[1]{Eq.~(\ref{#1})}
\newcommand{\Eq}[1]{Equation~(\ref{#1})}
\newcommand{\vp}[2]{[\mathbf{#1} \times \mathbf{#2}]}


\title{ Superfluid spin transport  in ferro- and antiferromagnets}

\author{E.  B. Sonin}
 \affiliation{Racah Institute of Physics, Hebrew University of
Jerusalem, Givat Ram, Jerusalem 91904, Israel}

\date{\today}

\begin{abstract}
This paper focuses on spin superfluid transport, observation of which was recently reported in antiferromagnet Cr$_2$O$_3$ [Yuan {\em et al.}, Sci. Adv. {\bf 4}, eaat1098 (2018)]. This paper analyzes the role of dissipation in transformation of spin current injected with incoherent magnons to a superfluid spin current near the interface where spin is injected.  The Gilbert damping parameter in the Landau--Lifshitz--Gilbert theory does  not describe dissipation properly, and the dissipation  parameters are calculated from the Boltzmann equation for magnons scattered by defects. The two-fluid theory is developed similar to the two-fluid theory for superfluids. This theory shows that the influence  of temperature variation in bulk on the superfluid spin transport (bulk Seebeck effect) is weak at low temperatures. The scenario that the results of Yuan {\em et al.} are connected with the Seebeck effect at the interface between the spin detector and the sample is also discussed.

The Landau criterion for an antiferromagnet put  in a magnetic field is derived from the spectrum of collective spin modes. The  Landau instability starts in the gapped mode earlier than in the Goldstone gapless mode, in contrast to easy-plane ferromagnets where  the Goldstone mode becomes unstable. The structure of the magnetic vortex in the geometry of the experiment is determined. The vortex core has the skyrmion structure with finite magnetization component normal to  the magnetic field. This magnetization creates stray magnetic fields around the exit point of  the vortex line from the sample, which can be used for experimental detection of vortices.
\end{abstract}

\maketitle


\section{Introduction} \label{Intr}

The concept of spin superfluidity is based on the analogy of the equations of magnetodynamics  with the equations of superfluid hydrodynamics.\cite{HalHoh}.
The analogy led to the suggestion that in magnetically ordered media persistent spin currents are possible, which are able to transport spin on macroscopical distances without essential losses.\cite{ES-78b} 

The phenomenon of spin superfluidity has been discussed for several decades.\cite{ES-78b,ES-82,Bun,Adv,BunV,Tserk,Halp,Mac,Pokr,Son17,Duine,Hoefer,BratAF,Sp1} We define the term superfluidity in its original meaning known from the times of Kamerlingh Onnes and Kapitza: transport of some physical quantity (mass, charge, or spin) over {\em macroscopical} distances without essential dissipation. This requires a constant or slowly varying phase gradient at macroscopic scale  with the total phase variation along the macroscopic sample equal to $2\pi$ multiplied by a very large number.  Spin superfluidity assumes the existence of spin current  proportional to the gradient of the phase (spin supercurrent). In magnetically ordered media the phase is an angle of rotation in spin space around  some axis (further in the paper the axis $z$). In contrast to the dissipative spin-diffusion current proportional to the gradient of spin density, the spin supercurrent is not accompanied by dissipation. 

Spin superfluidity require special topology of the order parameter space. This topology is realized at the presence of the easy-plane magnetic anisotropy, which confines the magnetization of the ferromagnet or sublattice magnetizations of the antiferromagnet in an easy plane. In this case one may expect that the current state is stable with respect to phase slips, which lead to relaxation of the supercurrent. In the phase slip event a vortex with $2\pi$ phase variation around it crosses streamlines of the supercurrent decreasing the total phase variation across streamlines by $2\pi$. The concept of the phase slip was introduced by \citet{And6} for superfluid $^4$He and later was used in studying spin superfluidity.\cite{ES-78b,ES-82}

Phase slips are suppressed by energetic barriers for vortex expansion. But these barriers  disappear when phase gradients reach critical values determined by the Landau criterion. The physical meaning of the Landau criterion is straightforward: the current state becomes unstable when there are elementary excitations with negative energy. So,  to check the Landau criterion one must know the full spectrum of collective modes.

 Sometimes any presence of spin current proportional to the phase gradient is considered as a manifestation of spin superfluidity.\cite{BunLvo,spinY} However, spin current proportional to the spin phase gradient  is ubiquitous and exists in any spin wave or domain wall, also in the ground state of disordered magnetic media.  In all these cases the total  variation of the phase  is smaller, or on the order of $\pi$.  Connecting these cases with spin superfluidity makes this phenomenon trivial and  already  observed  in old experiments on spin waves in the middle of  the 20th Century.  One may call the supercurrent  produced by the total phase variation of the order or less than $2\pi$  {\em microscopical} supercurrent, in contrast to persistent {\em macroscopical} supercurrents able to transport spin over  macroscopical distances. 

The analogy with usual superfluids is exact only if the spin space is invariant with respect to spin rotation around the hard axis normal to the easy plane. Then there is the conservation law for the spin component along the hard axis. In reality  this invariance is broken by in-plane anisotropy. But this anisotropy is usually weak, because it originates  from the spin-orbit interaction, which is relativistically small  compared to the exchange interaction, i.e.,  inversely  proportional to the speed of light.\cite{LLelDy} Macroscopical spin supercurrents are still possible if the energy of supercurrents exceeds the in-plane anisotropy energy. Thus, one cannot observe macroscopical spin supercurrents not only at large currents as in usual superfluids, but also at small currents.\cite{ES-78b}

From the time when the concept of spin superfluidity (in our definition of this term) was suggested\cite{ES-78b}, it was debated about whether the superfluid spin current is a ``real'' transport current. As a response to these concerns, in Ref~\onlinecite{ES-78b}  a {\em Gedanken} (at that time) experiment for demonstration of reality of superfluid spin transport was proposed. The spin is injected to one side of a magnetically ordered layer of thickness $d$ and spin accumulation is checked at another side. If the layer is not spin-superfluid, then the spin is transported by spin diffusion.  The spin current and the spin density exponentially decay at the distance of the spin diffusion length, and the density of  spin accumulated at the other side decreases exponentially  with growing distance $d$. However, if the conditions for spin superfluidity are realized in the layer, then the superfluid spin current decays much slower, and  the  accumulated spin density at the side opposite to the side where the spin is injected is inversely proportional to $d+C$, where $C$ is some constant.  

The interest to long-distance spin transport, especially to spin superfluid transport, revived recently. \citet{Tserk} carried out a microscopic analysis of injection of spin  to and ejection of spin out of the spin-superfluid medium in an easy-plane ferromagnet justifying the aforementioned scheme of superfluid spin transport. \citet{Halp} extended this analysis to easy-plane antiferromagnets.  Finally \citet{WeiH} were able to realize the suggested experiment  in antiferromagnetic Cr$_2$O$_3$  observing spin accumulation inversely proportional to the distance from the interface where spin was injected into  Cr$_2$O$_3$. 

Previously \citet{flux} reported  evidence of spin  superfluidity in the $B$ phase of superfluid $^3$He. They  detected  phase slips in a channel with superfluid spin current  close to its critical value. It was important evidence that persistent spin currents are possible.  But real long-distance transportation of spin by these currents was not demonstrated.  Moreover, it is impossible to do in the nonequilibrium magnon Bose--Einstein condensate, which was realized in the $B$ phase of $^3$He superfluid\cite{BunV} and in yttrium-iron-garnet magnetic films.\cite{Dem6} The nonequilibrium magnon Bose--Einstein condensate  requires  pumping of spin in the whole bulk for its existence. In the geometry of the aforementioned spin transport experiment this would mean that spin is permanently pumped not only by a distant injector but also all the way up the place where its accumulation is probed. Thus, the spin detector measures  not only spin coming from a distant injector but also  spin pumped close to the detector. Therefore, the experiment does not prove the existence of long-distance spin superfluid transport. There were also reports on experimental detection of spin superfluidity in magnetically ordered solids\cite{BunLvo,spinY}, but they addressed microscopical  spin supercurrent.\cite{SonConf}  As explained above, ``superfluidity'' connected with such currents was well proved by numerous old experiments on spin waves and does not need new experimental confirmations. The work of \citet{WeiH} was the first report on long-distance superfluid spin transport with spin accumulation  decreasing with  distance from the injector as expected from the theory.   Long distance superfluid spin transport was also recently reported in a graphene quantum antiferromagnet.\cite{Lau18}

The experiment on superfluid spin transport\cite{WeiH} has put to rest another old dispute about the spin superfluidity concept. At studying spin superfluidity in the $B$ phase of superfluid $^3$He, it was believed\cite{Bun} that spin superfluidity is possible only if there are mobile carriers of spin and a counterflow
 of carriers with opposite spins transports spin. If so, then spin superfluidity is impossible in insulators. Moreover, \citet{Niu} argued that it is  a critical  flaw of spin-current definition if it predicts spin currents in insulators. Since Cr$_2$O$_3$ is an insulator the experiment of \citet{WeiH} rules out this presumption.
 
 Boosted by the superfluid spin transport experiment\cite{WeiH} this paper addresses some  issues deserving further investigation. It is especially needed because \citet{Lebrun} made an experiment  in an antiferromagnetic iron oxide similar to that of \citet{WeiH} and observed similar dependence of spin accumulation on the distance from the injector.  However, \citet{Lebrun} explain it not by spin transport from the distant injector  but by  the Seebeck effect at the detector, which is warmed by the heat flow from the injector.  We shall compare these two interpretations in Sec.~\ref{DS}. 

We analyzed the role of dissipation in the superfluid spin transport. A widely used approach to address dissipation in magnetically ordered solids is  the Landau--Lifshitz--Gilbert (LLG) theory with the Gilbert damping parameter.  But we came to the conclusion that the Gilbert damping does not provide a proper description of dissipation processes in easy-plane ferromagnets. The Gilbert damping is described by a single parameter, which scales {\em all} dissipation processes independently from whether they do violate the spin conservation  law, or do not.  Meanwhile, the processes violating the spin conservation law, the Bloch spin relaxation in particular, originate from spin-orbit interaction and must be relativistically small as explained above. This requires the presence of a small factor in the intensity of the Bloch spin relaxation, which is absent in the Gilbert damping approach. So we determined the dissipation parameters from the Boltzmann equation for magnons scattered by defects. Dissipation is possible only in the presence of thermal magnons, and we developed the two-fluid theory for easy-plane ferromagnets similar to that in superfluid hydrodynamics for the clamped regime, when the gas of quasiparticles cannot freely drift without dissipation in the laboratory frame.

As mentioned above,  to check the Landau criterion for superfluidity, one must calculate the spectrum of collective modes and check whether some modes have negative energies. The Landau critical gradient is determined by easy-plane crystal anisotropy and was known qualitatively both for ferro- and antiferromagnets long ago.\cite{ES-78b} For easy-plane ferromagnets the Landau critical gradient was recently determined quantitatively from the spin-wave spectrum  in the analysis of ferromagnetic spin-1 BEC of cold atoms.\cite{Sp1} But Cr$_2$O$_3$, which was investigated in the experiment,\cite{WeiH}  has no crystal easy-plane anisotropy, and an ``easy plane'' necessary for spin superfluidity is produced by an external magnetic field. The magnetic field should exceed the spin-flop field, above which magnetizations of sublattices in antiferromagnet are kept in a plane normal to the magnetic field.  We analyze the magnon spectrum in the spin current states in this situation. The analysis has shown that the Landau critical gradient is determined by the gapped mode, 
but not by the Goldstone gapless mode as in the cases of easy-plane ferromagnets.

Within the two-fluid theory the role of spatial temperature variation was investigated. This variation produces the bulk Seebeck effect. But the effect is weak because it is proportional not to the temperature gradient, but to a higher (third) spatial derivative of the temperature. 

The transient processes near the interface through which spin is injected were also discussed. Conversion from spin current of  incoherent thermal magnons to coherent (superfluid) spin transport is among these processes. The width of the transient layer (healing length), where formation of the superfluid spin current occurs, can be determined by different scales at different condition. But at low temperatures it is apparently not less than the magnon mean-free-path.   

In reality the decay of superfluid currents starts at values less than the Landau critical value via phase slips produced by magnetic vortices. The difference in the spectrum of collective modes in ferro- and antiferromagnets leads to the difference in the structure of magnetic vortices. In the past magnetic vortices were investigated mostly in ferromagnets (see Ref.~\onlinecite{Sp1}  and references therein). The present work analyzes a vortex in an antiferromagnet. The vortex core has a structure of skyrmion with sublattice magnetizations deviated from the direction normal to the magnetic field. At the same time inside the core the total  magnetization has a component normal to the magnetic field. In the geometry of the Cr$_2$O$_3$ experiment this transverse magnetization creates surface magnetic charges at the point of the exit of the vortex line from the sample. Dipole stray magnetic fields produced by these charges hopefully can be used for detection of magnetic vortices experimentally.

Section \ref{SpTr} reminds the phenomenological model of Ref.~\onlinecite{ES-78b} describing the spin diffusion and superfluid spin transport. 
 Section~\ref{CMF} reproduces the derivation of the spectrum of the collective spin mode and the Landau criterion in a spin current state of an easy-plane ferromagnet known before\cite{Sp1}. This is necessary for comparison with the spectrum of the collective spin modes and the Landau criterion in a spin current state of an easy-plane antiferromagnet derived in Sec.~\ref{CMA}. Thus, Sec.~\ref{CMF}, as well as Sec.~\ref{SpTr}, do not contain new results, but were added to the paper to make it self-sufficient and more readable. In Sec.~\ref{Boltz} we address two-fluid effects and dissipation parameters  (spin diffusion and second viscosity coefficients) deriving them from the Boltzmann equation for magnons. The section also estimates the bulk Seebeck effect and shows that it is weak. Section \ref{Heal} analyzes the transient layer near the interface through which spin is injected and where the bulk superfluid spin current is formed. Various scales determining the width of this layer (healing length) are discussed.  In Sec.~\ref{AFV} the skyrmion structure of the magnetic vortex in an antiferromagnets is investigated.   The concluding Sec.~\ref{DS} summarizes the results of the work and presents some numerical estimations for the antiferromagnetic Cr$_2$O$_3$ investigated in the experiment. The Appendix analyzes dissipation in the LLG theory with the Gilbert damping. It is argued that this theory predicts dissipation coefficients incompatible with the spin conservation law.

\section{Superfluid spin transport vs spin diffusion} \label{SpTr} 

Here we remind the simple phenomenological model of spin transport suggested in Ref.~\onlinecite{ES-78b} (see also more recent Refs.~\onlinecite{Adv,Tserk,Halp}). The equations of magnetodynamics are
     \be
{dM_z \over dt}=-\bm \nabla \cdot \bm J- {M'_z \over T_1},
      \ee{EmB}
 \be
{d\varphi \over dt}=- {\gamma  M'_z\over \chi}+\zeta \nabla^2\varphi.
    \ee{EpB}
Here $\chi$ is the magnetic susceptibility along the axis $z$,  $\varphi$ is the angle of rotation  (spin phase) in the spin space around the axis $z$, and $M'_z =M_z-\chi H$ is a nonequilibrium part of the magnetization density along the magnetic field $H$ parallel to the axis $z$. The time $T_1$ is the Bloch time of the longitudinal spin relaxation. The  term $\propto \nabla^2\varphi$ in \eq{EpB} is an analog of the second viscosity in superfluid hydrodynamic.\cite{Kha71,EBS}  The magnetization density $M_z$ and the magnetization current $\bm J$ differ   from the spin density and the spin current by sign and by the gyromagnetic factor $\gamma$. Nevertheless, we shall call the current $\bm J$ the spin current  to stress its connection with spin transport. The total spin current $\bm J =\bm J_s +\bm J_d$ consists of the superfluid spin current
  \be
 \bm  J_s = {\cal A} \bm\nabla \varphi,
       \ee{curS}
and the spin diffusion current  
   \be
 \bm  J_d =- D\bm\nabla M_z.
       \ee{Jd}
 The pair of the hydrodynamical variables $(M_z,\varphi)$ is a pair of conjugate Hamiltonian variables analogous to the pair ``particle density--superfluid phase'' in superfluid hydrodynamics.\cite{HalHoh}

\begin{figure}[t]
\includegraphics[width=.5\textwidth]{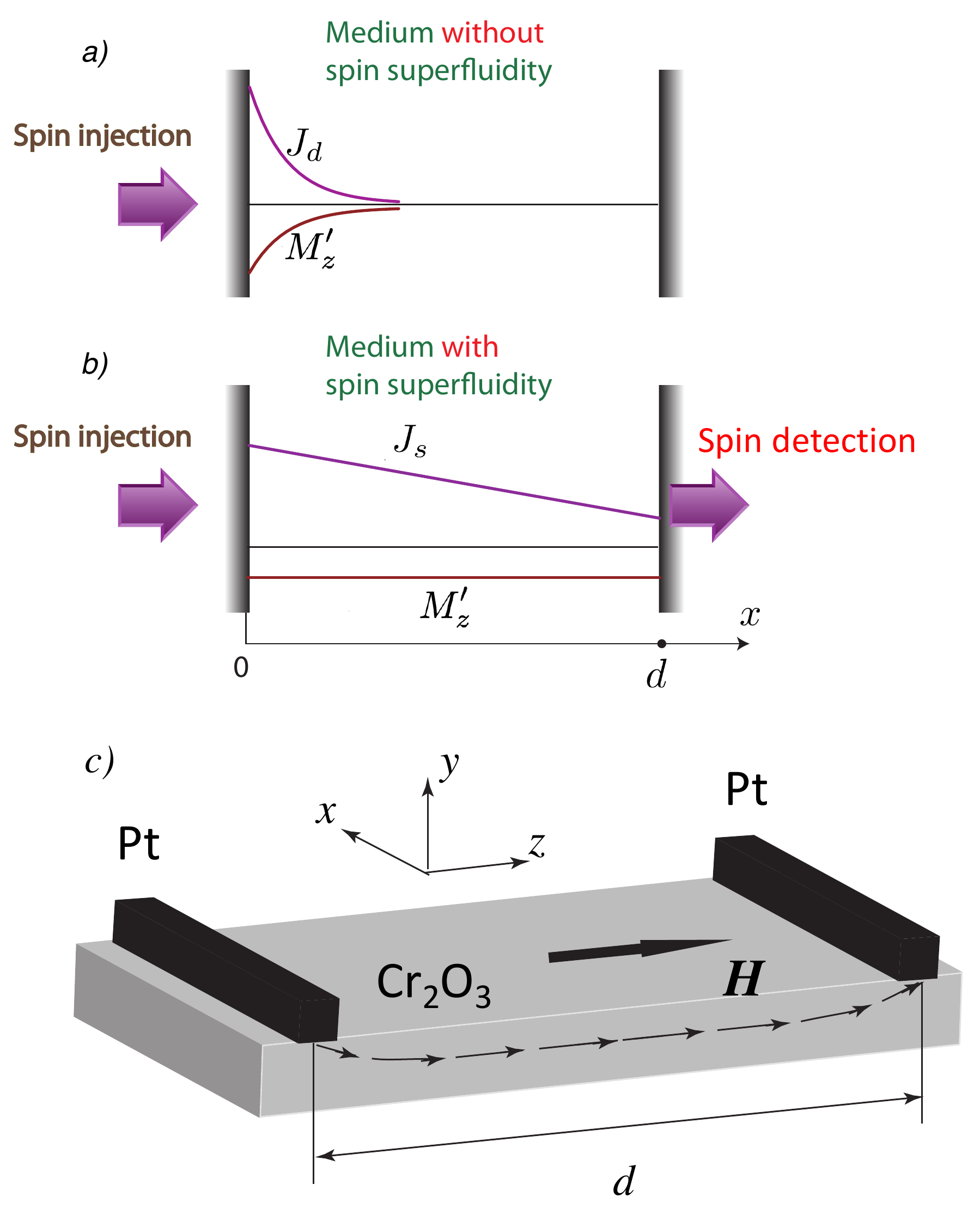}
\caption[]{ Long distance spin transport. (a) Spin injection to a spin-nonsuperfluid medium. (b) Spin injection to a  spin-superfluid medium. (c) Geometry of the experiment by \citet{WeiH}.
Spin is  injected from the left Pt wire and flows along the Cr$_2$O$_3$ film  to the right Pt wire, which serves as a detector. The arrowed dashed line shows a spin-current streamline. In contrast to  (a) and  (b), the spin current is directed along the same axis $z$ as a magnetization parallel to the external magnetic field $\bm H$. }
\label{f1}
\end{figure}

There are two kinds of spin transport illustrated in Fig.~\ref{f1}. In the absence of spin superfluidity (${\cal A}=0$) there is no superfluid current. \Eq{EpB} is not relevant, and \eq{EmB} describes pure spin diffusion [Fig.~\ref{f1}(a)].  Its solution, with the boundary condition that the spin current $J_0$ is injected at  the interface $x=0$, is
\be
J=J_d=J_0e^{-x/L_d}, ~~M'_z=J_0\sqrt{T_1\over D}  e^{-x/L_d},
   \ee{}
where 
\be
L_d=\sqrt{DT_1} 
   \ee{}
is the spin-diffusion length. Thus the effect of spin injection exponentially decays at the scale of the spin-diffusion length. 

However, if spin superfluidity is possible (${\cal A} \neq 0$), the spin precession equation (\ref{EpB}) becomes relevant. As a result of it, in a stationary state the magnetization $ M'_z$ cannot vary in space (Fig.~\ref{f1}b)  since according to \eq{EpB} the gradient $\bm\nabla M'_z$ is accompanied by the linear in time growth of the gradient $\bm\nabla \varphi$. The requirement of constant in space magnetization $M_z$  is similar to the requirement of constant in space chemical potential in superfluids, or the electrochemical potential in superconductors.  As a consequence of this requirement, spin diffusion current is impossible in the bulk since  it is simply ``short-circuited'' by the superfluid spin current. Only in AC processes the oscillating spin injection  can produce an oscillating bulk spin diffusion current coexisting with an oscillating superfluid spin current.

In the superfluid spin transport the spin current can reach the other boundary opposite to the boundary where spin is injected. We locate it at the plane $x=d$. As a boundary condition at $x=d$, one can use a phenomenological relation connecting the spin current with the magnetization:
$J_s(d) = M'_z v_d$, where $v_d$ is a phenomenological constant. This boundary condition was derived from the microscopic theory by \citet{Tserk}. Together with the boundary condition $J_s(0) = J_0$ at $x=0$ this yields the solution of Eqs.~(\ref{EmB}) and (\ref{EpB}):
 \be
M'_z= { T_1 \over d+v_dT_1}J_0,~~J_s(x) = J_0 \left(1-{x\over d+v_d  T_1} \right).
     \ee{}
Thus, the spin accumulated at large distance $d$ from the spin injector slowly decreases as the inverse distance $1/d$ [Fig.~\ref{f1}(b)], in contrast to the exponential decay $\propto e^{-d/L_d}$ in the spin diffusion transport [Fig.~\ref{f1}(a)].

In Figs.~\ref{f1}(a) and \ref{f1}(b) the spin flows along the axis $x$, while the magnetization and the magnetic field are directed along the axis $z$. In the geometry of the experiment of \citet{WeiH} the spin flows along the magnetization axis $z$ parallel to the magnetic field. This geometry is shown in Fig.~\ref{f1}c. The difference between two geometries is not essential if spin-orbit coupling is ignored. In this section we chose the geometry with different directions of the spin current and the magnetization in order to stress the possibility of the independent choice of axes in the spin and the configurational spaces. But in Sec.~\ref{AFV} addressing a vortex in an antiferromagnet we shall switch to the geometry of the experiment because in this case the difference between geometries is important.

Without dissipation-connected terms, the phenomenological theory of this section directly follows from  the LLG theory.  For  ferromagnets  the LLG equation  is
\be
{d\bm M\over dt}=\gamma \left[ \bm H_{eff}  \times \bm M\right],
       \ee{LLGf}
where   
\be
\bm H_{eff} =-{\delta {\cal H}\over \delta \bm M}= -{ \partial {\cal H}\over \partial \bm M}+\nabla_j{\partial {\cal H}\over \partial \nabla_j\bm M} 
   \ee{funDf}     
is the effective field determined by the functional derivative of the Hamiltonian $\cal H$.  For a  ferromagnet with uniaxial anisotropy the Hamiltonian is 
\bem
{\cal H}= {GM_z^2\over 2} +A \nabla_i \bm M\cdot \nabla_i \bm M - M _z H.
  \eem{}
Here $H$ is an external constant magnetic field parallel to the axis $z$, and the exchange constant  $A$ determines stiffness with respect to deformations of the magnetization field. In the case of easy-plane anisotropy the anisotropy parameter $G$ is positive and coincides  with the inverse susceptibility: $G=1/\chi$. 

Since the absolute value $M$ of the magnetization is a constant, one can describe  the 3D magnetization vector $\bm M$ only by two Hamiltonian conjugate variables: the magnetization $z$ component $M_z$ and the angle $\varphi$ of rotation around the $z$ axis. Then the LLG theory yields two equations 
\be
\dot M_z =-\bm\nabla\cdot \bm J_s ,
   \ee{canMf}
\be
 \dot\varphi  =- \gamma \mu , 
    \ee{canPf}
with the Hamiltonian in new variables
\be
{\cal H}={M_z^2\over 2\chi } +{A M_\perp^2  \nabla\varphi^2\over 2} +{A M^2  (\bm \nabla M_z)^2\over 2 M_\perp^2}-M_zH. 
    \ee{HamF}
Here $M_\perp =\sqrt{M^2-M_z^2}$, and the spin ``chemical potential'' and  the superfluid spin current  are
\be
\mu={\delta {\cal H}\over \delta M_z}={ \partial {\cal H}\over \partial M_z}-\nabla_j{\partial {\cal H}\over \partial \nabla_j M_z},~~\bm J_s = \gamma {\partial {\cal H}\over  \partial \bm \nabla \varphi}. 
    \ee{}

After substitution of explicit  expressions for functional derivatives of the Hamiltonian (\ref{HamF}) the equations become
\be 
{\dot M_z\over \gamma}=-\bm\nabla\cdot(AM_\perp^2 \bm \nabla \varphi),
   \ee{eqWd1}
\bem
{\dot \varphi\over \gamma}
=-M_z\left[{1\over \chi} -A(\bm\nabla\varphi)^2-{A M^2 (\bm \nabla M_z)^2\over M_\perp^4}\right]
\nonumber \\
+ {A M^2  \over  M_\perp^2}\bm\nabla^2 M_z+H.
       \eem{eqWd}
The equations (\ref{EmB}) and (\ref{EpB}) without dissipation terms follow from Eqs.~(\ref{eqWd1}) and  (\ref{eqWd}) after linearization with respect to small gradients $\bm\nabla \varphi$ and nonequilibrium magnetization $M'_z =M_z -\chi H$ and ignoring the dependence of the spin chemical potential $\mu$ on $\bm \nabla M_z$. Then  ${\cal A}=\gamma A M_\perp^2$,   and $M_\perp$  is determined by its value $\sqrt{M^2-\chi^2 H^2}$ in the equilibrium. 

\section{Collective modes and the Landau criterion  in easy-plane ferromagnets} \label{CMF}

To check the Landau criterion one should know the spectrum of collective modes. In  an easy-plane ferromagnet   the collective modes (spin waves) are determined by Eqs.~(\ref{eqWd1})  and (\ref{eqWd}) linearized with respect to weak perturbations of stationary states. Further the angle variable $\theta$ will be introduced instead of the variable $M_z=M\sin\theta$.
Let us consider  a current state with constant gradient $\bm K=\bm  \nabla \varphi$ and constant magnetization 
\be
M_z=M\sin\theta={\chi H\over 1-\chi AK^2}.
      \ee{} 
To derive the spectrum of collective modes, we consider weak perturbations $\Theta$ and $\Phi$  of this state:  $\theta  \to \theta + \Theta$,   $\varphi  \to \varphi + \Phi$. Equations~(\ref{eqWd1}) and (\ref{eqWd}) after linearization are:
\bem
\dot \Theta- 2\gamma M_z A \bm K \cdot\bm\nabla  \Theta=- \gamma A M \cos\theta  \nabla^2\Phi   ,
\nonumber \\
\dot \Phi- 2\gamma M_z A \bm K \cdot\bm\nabla \Phi= 
\nonumber \\
-  { \gamma M\cos\theta\over \chi} \left(1 - \chi A K^2\right) \Theta+\gamma AM\cos\theta  \nabla^2\Theta.
      \eem{eqFer}
For plane waves $\propto e^{i\bm k\cdot \bm r-i\omega t}$ these equations describe the gapless Goldstone mode with the spectrum:\cite{Hoefer,Sp1}
\be
(\omega +\bm w \cdot \bm k)^2 = \tilde c_s^2 k^2.
    \ee{spF}
Here
\be
\tilde c_s = \sqrt{ \chi \over \tilde\chi} c_s,
      \ee{}
\be
\tilde \chi =\frac{\chi }{1 - \chi A \left(K^2 -{M^2 k^2\over M_\perp^2}\right)},
      \ee{chi}
and  
\be
c_s =\gamma M_\perp \sqrt{A\over \chi}
    \ee{}
is the spin-wave velocity in the ground state  without any spin current. In this state the spectrum becomes
\be
\omega =c_s k\sqrt{1 +\chi A {M^2 k^2\over M_\perp^2}}.
    \ee{fr0}
The velocity
\be
\bm w =2\gamma M_z A \bm K,
     \ee{}    
can be called Doppler velocity because its effect  on the mode frequency is similar to the effect of the mass velocity on the mode frequency in a Galilean invariant fluid (Doppler effect). But our system is not Galilean invariant,\cite{Hoefer} and the gradient $K$ is present also in the right-hand side of the dispersion relation (\ref{spF}). 

In the long-wavelength hydrodynamical limit magnons have the sound-like spectrum linear in $k$. Quadratic corrections $\propto k^2$ become important at $k \sim M_\perp/M\sqrt{ \chi A}$ [see \eq{fr0}]. These corrections emerge from the terms in the Hamiltonian, which depend on $\bm \nabla M_z$. So the hydrodynamical approach is valid at scales exceeding 
\be
\xi_0={M\over M_\perp}\sqrt{ \chi A},
    \ee{xi0}
which can be called the coherence length, in analogy with the coherence length in the Gross--Pitaevskii theory for BEC. Also in analogy with BEC, the coherence length diverges at $M_\perp \to 0$, i.e., at the second-order phase transition from the easy-plane to the easy-axis anisotropy. 
 The same scale determines the Landau critical gradient and the vortex core radius. Telling about hydrodynamics we bear in mind hydrodynamics of a perfect fluid without dissipation. Later in this paper we shall discuss hydrodynamics with dissipation. In this case the condition $k\ll  1/\xi_0$ is not sufficient, and an additional restriction on using hydrodynamics is determined by the mean-free path of magnons.

According to the Landau criterion, the current state becomes unstable at small $ k$ when $ \bm k$ is parallel to $\bm w$ and the frequency $\omega$ becomes negative. This happens at the gradient  $K$ equal to the Landau critical gradient
\be
K_c={M_\perp \over\sqrt{ 4M^2 - 3M_\perp}} {1\over \sqrt{\chi A}}\sim {1\over \xi_0}.
     \ee{KcFg}
Spin superfluidity becomes impossible at the phase transition to the easy-axis anisotropy  ($M_\perp=0$). 
In the opposite limit of small $M_z\ll M$ the pseudo-Doppler effect is not important, and the Landau critical gradient $K_c$ is determined from the condition that the spin-wave velocity $\tilde c_s$ vanishes at small  $k $:
\be
K_c= {1\over \sqrt{\chi A}}= {\gamma  M \over \chi c_s}.
     \ee{KcF}

Expanding the Hamiltonian (\ref{HamF}) with respect to weak perturbations $\Theta$ and $\Phi$ up to the second order one obtains the energy of the spin wave mode per unit volume, 
\be
E_{sw} ={M_\perp \omega(\bm k)\over \gamma  \sqrt{\tilde \chi A } k} |\Theta_k|^2,
     \ee{}
where  $ |\Theta_k|^2  $ is the squared perturbation of the angle $\theta$ with the wave vector $\bm k$ averaged over the wave period.

In the quantum theory the energy density $E_{sw}$ corresponds to the magnon density
\be
 {n(\bm k)\over V}= {E_{sw} \over \hbar \omega(\bm k)}={M_\perp  |\Theta_k|^2 \over  \hbar \gamma  \sqrt{\tilde \chi A } k} ,
     \ee{}
where $n(\bm k)$ is the number of magnons in the plane-wave mode with the wave vector $\bm k$ and $V$ is the volume of the sample.   Summing over the whole $\bm k$ space, the averaged squared perturbation is 
\be
\langle \Theta ^2 \rangle=\sum_{\bm k}|\Theta_k|^2 ={ \hbar \gamma  \sqrt{ A } \over  M_\perp} \int \sqrt{\tilde \chi  } n(\bm k)k{d_3\bm k\over (2\pi)^3}.
      \ee{int}

Further we proceed within the hydrodynamical approach neglecting quadratic corrections to the spectrum. There are quadratic in spin-wave amplitudes corrections to the spin superfluid current and to the spin chemical potential:
\bem
\left. \bm J_s \right|_{sw}=-\gamma M_\perp A(M_\perp \langle \Theta ^2 \rangle \bm K+2 M_z \langle \Theta \bm \nabla \Phi \rangle ) ,
    \eem{}
\bem
\left. \mu \right|_{sw}
=- A(M_z \langle(\nabla\Phi)^2 \rangle+2M_\perp \bm K \cdot \langle\Theta \bm \nabla\Phi\rangle ).
    \eem{}
Using \eq{int} and  the relation 
\be
\bm \nabla \Phi= {\Theta\over \sqrt{\chi A}}{\bm k\over k},
    \ee{}
which follows from the equations of motion (\ref{eqFer}), one obtains: 
\be
\left. \bm J_s \right|_{sw}=- {\chi^2   \hbar c_s^3  \over  \gamma M_\perp^2} \int n(\bm k) \left(\bm K+{2\gamma M_z   \over   \chi   c_s}{\bm k\over k} \right)k{d_3\bm k\over (2\pi)^3} ,
    \ee{cuP}
\be
\left.\mu \right|_{sw}
= -  { \chi \hbar c_s^2 \over \gamma  M_\perp^2 } \int n(\bm k)\left( {\gamma  M_z\over  \chi  c_s }+{2\bm K \cdot\bm k\over k} \right)k{d_3\bm k\over (2\pi)^3} .
    \ee{precP}

\section{Collective modes and the Landau criterion  in antiferromagnets} \label{CMA}

 For ferromagnetic state of localized spins the derivation of the LLG theory from the microscopic Heisenberg model was straightforward.\cite{LLstPh2}  The quantum theory of the antiferromagnetic state even for the  simplest case of a two-sublattice antiferromagnet, which was widely used for Cr$_2$O$_3$, is more difficult.  This is because the state with constant magnetizations of two sublattices is not a well defined quantum-mechanical eigenstate.\cite{LLstPh2} Nevertheless, long time ago it was widely accepted to ignore this complication and to describe the  long-wavelength dynamics by the LLG theory for two sublattices coupled via exchange interaction:\cite{Kittel51} 
\be
{d\bm M_i\over dt}=\gamma \left[\bm H_i  \times \bm M_i\right],
       \ee{LLG}
 where the subscript  $i=1,2$ points out to which sublattice the magnetization $\bm M_i$ belongs,   and
 \be
\bm H_i =-{\delta {\cal H}\over \delta \bm M_i}= - {\partial {\cal H}\over \partial \bm M_i}+\nabla_j{\partial {\cal H}\over \partial \nabla_j\bm M_i} 
   \ee{}     
is the effective field for  the $i$th  sublattice determined by the functional derivative of the Hamiltonian $\cal H$.  For an isotropic antiferromagnet  the Hamiltonian is 
\bem
{\cal H}= {\bm M_1 \cdot \bm M_2\over \chi} + { A (\nabla_i \bm M_1 \cdot \nabla_i \bm M_1+\nabla_i \bm M_2 \cdot \nabla_i \bm M_2)\over 2}
\nonumber \\
+A_{12} \nabla_j \bm M_1 \cdot \nabla_j \bm M_2-\bm H \cdot(\bm M_1+\bm M_2).~~~
  \eem{ham2}
In the uniform ground state without the magnetic field $\bm H$  the two magnetizations are antiparallel, $\bm M_2=-\bm M_1$, and the total magnetization $\bm M_1+\bm M_2$ vanishes. At  $\bm H\neq 0$ the sublattice magnetizations are canted, and in the uniform ground state  the total magnetization is parallel to $\bm H$: 
\be
 \bm m=\bm M_1+\bm M_2 =\chi \bm H.
     \ee{}
 The first term in the Hamiltonian (\ref{ham2}), which determines the susceptibility $\chi$,  originates from the exchange interaction between spins of two sublattices. This is the susceptibility normal to the staggered magnetization (antiferromagnetic vector) $\bm L=\bm M_1-\bm M_2$. Since in the LLG theory absolute values of magnetizations $\bm M_1$ and $\bm M_2$ are fixed the susceptibility parallel to $\bm L $ vanishes.
 
  \begin{figure}[b]
\includegraphics[width=.3\textwidth]{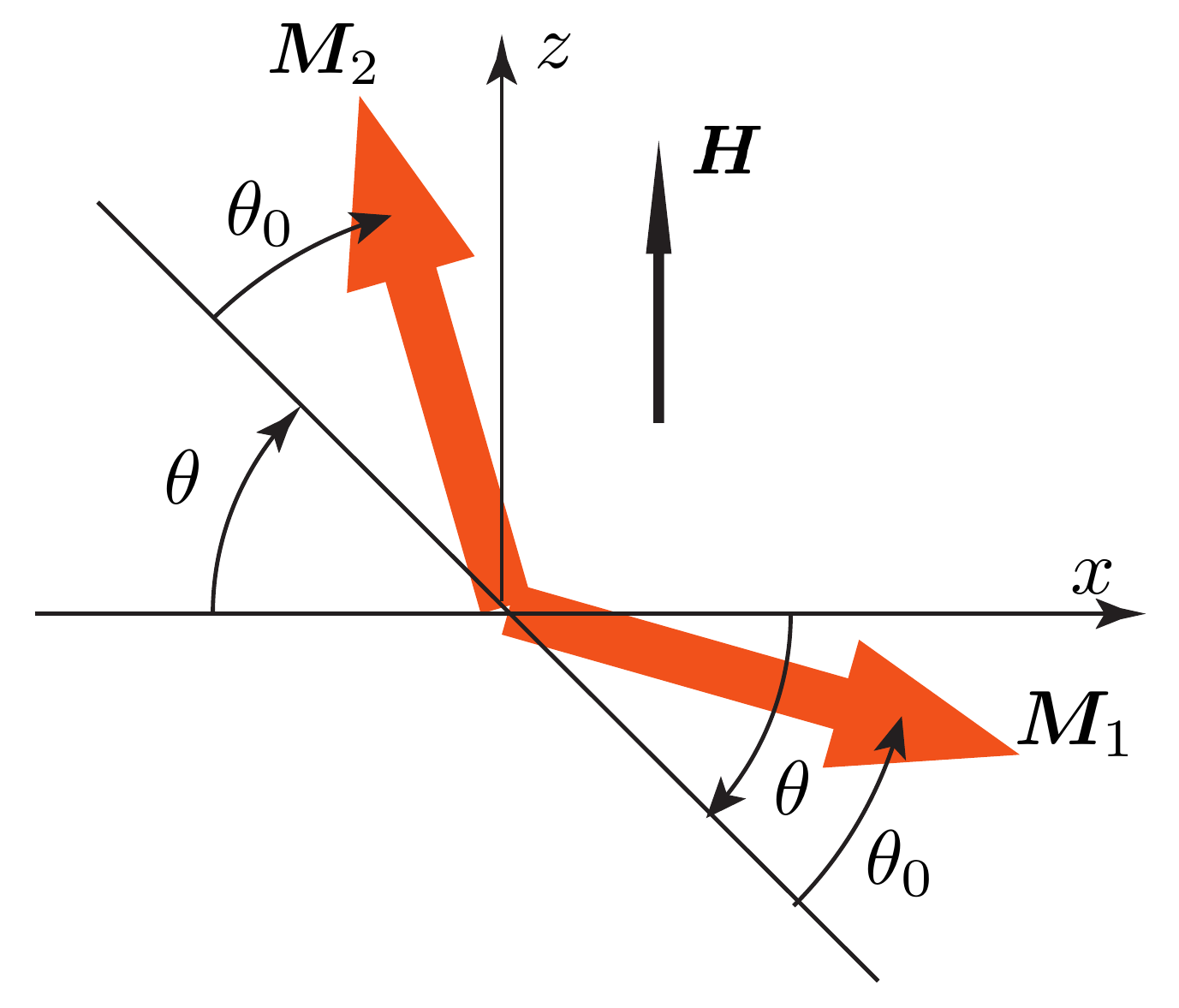}
\caption[]{ Angle variables $\theta$ and $\theta_0$  for the case when the both magnetizations are in the plane $xz$ ($\varphi_0 =\varphi=0$).}
\label{f2}
\end{figure}

In the uniform state only the uniform exchange energy $\propto 1/\chi$ and the Zeeman energy (the first and the last terms) are present in the Hamiltonian, which can be rewritten as
\be
{\cal H}= -{L^2-m^2 \over  4 \chi} -\bm H \cdot \bm m = -{M^2 \over   \chi}+{m^2 \over  2 \chi} - m H_m,
  \ee{ham2r}
where $H_m= (\bm H \cdot \bm m)/m$ is the projection of the magnetic field on the direction of the total magnetization $\bm m$. Minimizing the Hamiltonian with respect to the  absolute value of $\bm m$ (at it fixed direction, i.e., at fixed $H_m$) one obtains
\bem
{\cal H}=- {M^2 \over   \chi}-{\chi  H_m^2 \over  2 }=- {M^2 \over   \chi}-{\chi  H^2 \over  2 } +{\chi  H_L^2 \over  2},
  \eem{}
where $H_L =(\bm H \cdot \bm L)/L$  is the projection of the magnetic field on the staggered magnetization $\bm L$.  The first two terms are constant, while the last term plays the role of the easy-plane anisotropy energy confining $\bm L$ in the plane normal to $\bm H$. For $\bm H$ parallel to the axis $z$:
\be
E_a = {\chi  H^2 L_z^2 \over 2 L^2}= {\chi  H^2\sin\theta \over 2 }.
        \ee{Ea}
Here $\theta$ is the angle between the staggered magnetization $\bm L$ and the $xy$ plane (see  Fig.~\ref{f2}).  

We introduce the pairs of angle variables $\theta_i$, $\varphi_i$ determining directions of the sublattice magnetizations:
\bem
M_{ix} =M\cos\theta_i\cos \varphi_i,~~M_{iy} =M\cos\theta_i\sin \varphi_i,
\nonumber \\
M_{iz}=M\sin \theta_i.
       \eem{}
 The equations  of motion in the angle variables are
\bem
{\cos \theta_i\dot \theta_i\over \gamma}={1\over M}\left({\partial {\cal H}\over \partial \varphi_i}-\nabla{\partial {\cal H}\over \partial \nabla \varphi_i}\right),
\nonumber \\
{\cos \theta_i\dot\varphi_i \over \gamma} =- {1\over M}\left({\partial {\cal H}\over \partial \theta_i}-\nabla{\partial {\cal H}\over \partial \nabla \theta_i}\right).
    \eem{canV}
In the further analysis it is convenient to use other angle variables:
\bem
\theta_0={\pi +\theta_1-\theta_2\over 2},~~ \theta ={\pi-\theta_1-\theta_2\over 2},
\nonumber \\
\varphi_0={\varphi_1+\varphi_2\over 2},~~\varphi={\varphi_1-\varphi_2\over 2}.
    \eem{}
In these variables the Hamiltonian becomes
\begin{widetext}
\bem
{\cal H}=-{M^2\over \chi}( \cos 2\theta_0 \cos^2 \varphi - \cos 2\theta \sin^2 \varphi)-2H M\cos \theta\sin  \theta_0 
\nonumber \\
+AM^2[(1+\cos 2\theta_0 \cos 2\theta ){\nabla \varphi_0^2+\nabla \varphi^2\over 2} -\sin 2\theta_0 \sin 2\theta \bm \nabla \varphi_0\cdot \bm\nabla \varphi +\nabla \theta_0^2+\nabla \theta^2]
\nonumber \\
+A_{12} M^2\{(\cos 2\theta  \sin^2\varphi+\cos 2 \theta_0 \cos ^2\varphi)(\nabla \theta_0^2-\nabla \theta^2 )-{\cos 2\theta_0+ \cos  2\theta\over 2}\cos 2\varphi ( \nabla \varphi_0^2-\nabla \varphi^2)
\nonumber \\
-\sin2\varphi [ \sin 2\theta  (\bm \nabla \theta_0 \cdot \bm \nabla \varphi_0  +\bm \nabla \theta  \cdot \bm \nabla \varphi)+ \sin 2\theta_0 (\bm  \nabla \theta  \cdot \bm  \nabla \varphi_0+\bm \nabla \theta_0  \cdot \bm \nabla \varphi)] \} .
   \eem{H}
\end{widetext}
The polar angles $\theta$ for the staggered magnetization $\bm L$ and the canting angle $\theta_0$ are shown in Fig.~\ref{f2} for the case when the both magnetizations are in the plane $xz$ ($\varphi_0 =\varphi=0$). 

     In the uniform ground state $\theta=0$,  $\varphi=0$, $m_z =2M \sin\theta_0=\chi  H$, while the angle $\varphi_0$ is an arbitrary constant. Since we consider fields $H$ weak compared to the exchange field, $\theta_0$ is always small.
In the state with  constant  current $\bm K=\bm \nabla \varphi_0$ the magnetization along the magnetic field is
\be
 m_z = {\chi H \over 1-\chi A_- K^2/2},
   \ee{}
where $A_\pm=A\pm A_{12}$.

In a weakly perturbed current state  small but nonzero  $\theta$ and $\varphi$ appear. Also the angles $\theta_0$ and $\varphi_0$ differ from their values in the stationary current state:  $\theta_0 \to \theta_0+ \Theta$,   $\varphi _0\to\varphi_0+ \Phi$. Linearization of the nonlinear equations of motion with respect to  weak perturbations $\Theta$, $\Phi$, $\theta$, and $\varphi$ yields decoupled linear equations for two pairs of variables $(\Theta,\Phi)$ and $(\theta,\varphi)$:
\bem
{\dot \Theta\over \gamma}-A_-m_z  \bm K\cdot \bm \nabla \Theta=  -A_-M _\perp  \nabla^2 \Phi,   
\nonumber \\
{\dot\Phi \over \gamma}-A_-m_z \bm K\cdot \nabla \Phi
 =-  \left(1- {\chi A_- K^2\over 2}\right){2M_\perp\over \chi}\Theta
 \nonumber \\
 +{(A+A_{12} \cos 2\theta_0)\over \cos \theta_0} M   \nabla^2 \Theta,
      \eem{gapL}
\bem
{\dot \theta \over \gamma}- A_+m_z \ \bm K\cdot \bm \nabla \theta
 \nonumber \\
 = -{2 M_\perp\over \chi}\left(1+\chi A_{12} K^2\right) \varphi
+ A_+M_\perp \nabla^2 \varphi ,
\nonumber \\{\dot\varphi \over \gamma}- A_+m_z \cos \theta_0 \bm K\cdot \nabla \varphi 
\nonumber \\
= {m_z^2\over 2\chi M_\perp} (1+\chi A_{12}K^2)  \theta -A_-K^2M_\perp \theta  
\nonumber \\
-{A-A_{12} \cos 2 \theta_0\over \cos\theta_0}M \nabla^2\theta.
   \eem{gap}
For plane waves $\propto e^{i\bm k\cdot \bm r-i\omega t}$ \eq{gapL} describes the gapless Goldstone mode with the spectrum:
\bem
(\omega +\gamma m_z A_-  \bm K\cdot \bm k)^2
\nonumber \\
= c_s^2 \left[1 -{\chi A_-  K^2\over 2}+{\chi (A+A_{12} \cos 2\theta_0) k^2\over 2\cos^2 \theta_0}\right]k^2.
    \eem{}
Here
\be
c_s =\gamma M_\perp \sqrt{2 A_-\over \chi}
     \ee{}
is the spin-wave velocity in the ground state without spin current. Apart from quadratic corrections $k^2$ to the frequency, the gapless mode in an antiferromagnet does not differ from that in a ferromagnet, if one replaces in all expressions for the ferromagnet   $A$ by $A_- /2$ and the parameter $M$ by $2M$.

\Eq{gap} describes the gapped mode with the spectrum
\bem
(\omega +\gamma m_z A_+   \bm K\cdot \bm k)^2=\left( 1 +\chi A_{12} K^2+{\chi A_+ k^2\over 2} \right)
\nonumber \\
\times \left[ {(1+\chi A_{12}K^2) \gamma^2 m_z^2\over \chi^2}  -c_s^2K^2  
\right. \nonumber \\ \left. 
+{2 \gamma^2M^2(A-A_{12} \cos 2 \theta_0) k^2\over \chi} \right].~
    \eem{spG}
 Without spin current  and neglecting the term $\propto A_+k^2$ the spectrum is
\be
\omega=\sqrt{ {\gamma^2m_z ^2\over \chi^2}  + c_s^2k^2 }.
    \ee{sp0}
This spectrum determines a new correlation length
\be
\xi={M\over H}\sqrt{2A_-\over \chi} ={c_s\over \gamma H},
    \ee{xi}
which is connected with the easy-plane anisotropy energy (\ref{Ea}) and determines the wave vector $k = 1/\xi$ at which the gap and the $k$ dependent frequency become equal.

Applying the Landau criterion to the gapless mode one obtains  the critical gradient $\sqrt{2/\chi A_-}$  similar to the value (\ref{KcF}) obtained for a ferromagnet. But in contrast to a ferromagnet where the susceptibility $\chi$ is connected with weak anisotropy energy, in an antiferromagnet the susceptibility $\chi$ is determined by a much larger exchange energy and is rather small. As a result, in an antiferromagnet the gapless Goldstone mode  becomes unstable at the very high value of $K$. But at much lower values of $K$ the gapped mode becomes unstable. According to the spectrum (\ref{spG}), the gap in the spectrum vanishes at the critical gradient 
\be
K_c ={1\over \xi} ={\gamma H\over c_s}  ={\gamma m_z \over \chi c_s}.
   \ee{} 
  
\section{Two-fluid effects and dissipation from the Boltzmann equation for magnons} \label{Boltz}

Knowledge of the spectrum of collective modes allows to  derive the dynamical equations at finite temperatures taking into account the presence of thermal magnons. Further we follow the procedure of the derivation  of  the two-fluid hydrodynamics in  superfluids.\cite{Kha71} We address the hydrodynamical limit when all parameters ($M_z$, $\bm K$, $T$) of the system slowly vary in space and time. 

We shall focus on ferromagnets. The equilibrium  Planck distribution of magnons in a ferromagnet  with a small spin current $\propto \bm K$ is
\be
n_{\bm K} ={1\over e^{\hbar \omega(\bm k) /T}-1} \approx n_0(\omega_0)  - {2\chi c_s^2 M_z\over \gamma M_\perp^2}{\partial n_0(\omega_0)\over   \partial \omega_0} \bm K \cdot\bm k, 
    \ee{Pl} 
where $\omega_0 =c_sk$ and 
\be
n_0(\omega_0) = {1\over e^{\hbar \omega_0 /T}-1}
     \ee{}
is the Planck distribution in the state without spin current. 

In the theory of superfluidity the Plank distribution of phonons in general depends not only on density and superfluid velocity (analogs of our $M_z$ and $\bm K$) but also on the normal velocity, which characterizes a possible drift of the gas of quasiparticles with respect to the laboratory frame of coordinates. This drift is possible because of the Galilean invariance of superfluids. In  our case the Galilean invariance is broken by possible interaction of magnons with defects, and in the equilibrium the drift of the quasiparticle gas is impossible. The case  of broken Galilean invariance, when the normal velocity vanishes, was also investigated for superfluids in porous media or in very thin channels, when the Galilean invariance is broken by interaction with channel walls. It was called the clamped regime.\cite{atkins,beelen}

Substituting the Planck  distribution (\ref{Pl}) into Eqs.~(\ref{cuP}) and (\ref{precP}) one obtains the contribution of equilibrium magnons to the spin current and the spin chemical potential:
\bem
\left. \bm J_s \right|_{eq}=\gamma {\partial \Omega\over \partial \bm K}=-  {\pi^2\chi^2   T^4\over 30 \gamma M_\perp^2 \hbar^3 c_s } \bm K   \left(1+{ 16M_z^2   \over 3 M_\perp^2}  \right),
    \eem{cuE}
\be
\left. \mu  \right|_{eq}= {\partial \Omega\over \partial M_z }=   {\pi^2 M_z T^4\over 30 \hbar^3 c_s^3 M_\perp^2} ,
    \ee{precE}
where
\be   
\Omega = T\int \ln(1-e^{-\hbar \omega(\bm k) /T}) {d_3\bm k\over (2\pi)^3}.
     \ee{Omega}
is the thermodynamical potential for the magnon Bose-gas.  The contribution (\ref{cuE}) decreases the superfluid spin current at fixed phase gradient $\bm K$, similarly to the decrease of the mass superfluid current after replacing the total mass density by the lesser superfluid density. 

\citet{WeiH} used in their experiment very thin film at low temperature, when  de Broglie wavelength of magnons  exceeds film thickness, and it is useful  to give also the two-fluid corrections for a two-dimensional case. Repeating our calculations after replacing integrals $\int d_3\bm k/(2\pi)^3$ by integrals $W\int d_2\bm k/(2\pi)^2$, one obtains:
\be
\left. \bm J_s \right|_{eq}=-  {\zeta(3)\chi^2   T^3\over \pi W \gamma M_\perp^2 \hbar^2 } \bm K   \left(1+{ 6M_z^2   \over M_\perp^2}  \right),
    \ee{cuE2}
\be
\left. \mu \right|_{eq}=  {\zeta(3) M_z T^3\over \pi W \hbar^2 c_s^2 M_\perp^2} ,
    \ee{precE2}
where the value of the Riemann zeta function $\zeta(3)$ is 1.202 and $W$ is the film thickness.

The next step in derivation of the two-fluid theory at finite temperatures  is the analysis of dissipation. A widely used approach of studying dissipation in magnetically order systems is the LLG theory with the Gilbert damping term added. However, this approach is incompatible with the spin conservation law. This law, although being approximate, plays a key role in the problem  of spin superfluidity. Therefore, we derived dissipation parameters from the Boltzmann equation for magnons postponing discussion of  the LLG theory with the Gilbert damping to the Appendix.

Dissipation is connected with nonequilibrium corrections to the magnon distribution. At low temperatures the number of magnons is small, and magnon-magnon interaction is weak. Then the main source of dissipation is scattering of magnons by defects. 
The Boltzmann equation  with the collision term in the relaxation-time approximation is
\be 
\dot n +{\partial \omega \over \partial \bm k}\cdot  \bm \nabla n- \bm \nabla\omega  \cdot  {\partial n \over \partial \bm k}=- {n-n_{\bm K}\over \tau}.
      \ee{Boltz0}
If  parameters, which determine the magnon distribution function $n$, vary slowly in space and time  one can substitute the equilibrium Planck distribution $n_{\bm K}$ into the left-hand side of the Boltzmann equation (\ref{Boltz0}). This yields:
\be 
{\partial n_0\over \partial \omega} \dot \omega+{\partial n_0\over \partial T}\left( \dot T +{\partial \omega \over \partial \bm k}\cdot  \bm \nabla T\right)=- {n-n_0\over \tau},
      \ee{Boltz1}
We consider small gradients $\bm K$ when the difference between $n_{\bm K}$ and $n_0$ is not important. But weak dependence of $\omega$ on  $\bm K$ is important at calculation of $\dot \omega $. One can see that at the constant temperature $T$ in any stationary state the left-hand side vanishes, and there is no nonequilibrium correction to the magnon distribution. Correspondingly, there is no dissipation. This is one more illustration that stationary superfluid currents do not decay.   

In nonstationary cases time derivatives are determined by the equations of motions. The equations of motion for $M_z$ and $\bm K$ are not sufficient, and the equation of heat balance is needed for finding $\dot T$. In general the heat balance equation is rather complicated since it must take into account interaction of magnons with other subsystems, e. g., phonons. Instead of it we consider a simpler case, when  magnons are not important in the heat balance, i.e.,  the temperature does not depend on magnon processes. In other words we consider the isothermal regime when $\dot T=0$. But we allow slow temperature variation in space.

The temporal variation of the frequency $\omega$ emerges from slow temporal variation of $M_z$ and $\bm K$, and at small $\bm K$
\be
\dot\omega ={\partial \omega \over \partial M_z}\dot M_z +{\partial \omega \over \partial  \bm K} \bm{\dot K}=- {M_z \over M_\perp^2}\left(c_sk \dot M_z 
 + {2\chi c_s^2 \over \gamma }\bm k\cdot  \bm{\dot K}\right).
    \ee{}
The partial derivatives ${\partial \omega / \partial M_z}$ and ${\partial \omega / \partial \bm K} $ were determined from the spectrum (\ref{spF}), while the time derivatives of $M_z$ and $ \bm K$ were found from the linearized equations (\ref{eqWd1}) and (\ref{eqWd}) assuming that $\bm \nabla \varphi=\bm K$ is small and ignoring gradients of $M_z$ in the right-hand side of \eq{eqWd}, which are beyond  the hydrodynamical limit. Then
\be
\dot\omega = {M_z \over M_\perp^2}c_s^2 \left[{\chi  \over \gamma }c_s k \bm \nabla \cdot \bm K + 2 (\bm k\cdot  \bm \nabla) M_z\right].
    \ee{}
Eventually the nonequilibrium correction to the magnon distribution function is
\bem
n'=n-n_0 =- {M_z \over M_\perp^2}c_s\left[ {\chi c_s\over \gamma }k  \bm \nabla \cdot \bm K 
\right. \nonumber \\ \left.
+ 2(\bm k\cdot  \bm \nabla) M_z-{M_\perp^2 \over M_z T}(\bm k\cdot  \bm \nabla)T  \right]\tau {\partial n_0\over \partial k}
    \eem{}

Substituting $n'$ into Eqs.~(\ref{cuP}) and (\ref{precP}) one obtains dissipation terms in the spin current and the spin chemical potential: 
\be
 \bm J_d=-D\left(\bm \nabla M_z -{1\over 2T}{M_\perp^2\over M_z} \bm \nabla T  \right) ,
    \ee{cuD}
\be
\mu_d= -{\zeta\over \gamma}\bm \nabla \cdot \bm K ,
    \ee{precD}
where
\bem
D= - {2\chi\hbar c_s^3\over 3\pi^2  } { M_z^2\over M_\perp^4}\int \tau    {\partial n_0\over \partial k} k^4\,d k,
\nonumber \\
\zeta=  - {\chi\hbar c_s^3\over 2\pi^2  } { M_z^2\over M_\perp^4}\int \tau    {\partial n_0 \over \partial k} k^4\,d k.
    \eem{}
In addition to the spin diffusion current, the dissipative spin current $\bm J_d$ contains also the current proportional to the temperature gradient. This is the bulk Seebeck effect. Estimation of the integral in these expressions requires knowledge of possible dependence of the relaxation time $\tau$ on the energy. Under the assumption that $\tau$ is independent from the energy, 
\be
D= {8 \pi^2 \tau  \gamma^2 T^4   M_z^2\over 45 \hbar^3 c_s^3 M_\perp^2},
~~\zeta=   {2 \pi^2 \tau  \gamma^2 T^4   M_z^2\over 15 \hbar^3 c_s^3 M_\perp^2},
    \ee{alp}
or for the two-dimensional case,
\be
D =  {16\zeta(3) \tau  \gamma^2 T^3   M_z^2\over 3\pi W   \hbar^2 c_s^2 M_\perp^2},~~\zeta =  {4\zeta(3) \tau  \gamma^2 T^3   M_z^2\over \pi W   \hbar^2 c_s^2 M_\perp^2}.
    \ee{alp2}

Although in antiferromagnets the Landau critical gradient is connected with the gapped mode, at small phase gradients the gapless Goldstone mode has lesser energy, and at low temperatures most of magnons  belong to this mode. Since the Goldstone modes in ferromagnets and antiferromagnets are similar, our estimation of dissipation coefficients for ferromagnets is valid also for antiferromagnets after replacing $A$ by $A_- /2$ and  $M$ by $2M$.

The microscopic analysis of this section agrees with the following phenomenological equations similar to the hydrodynamical equations for superfluids in the clamped regime:
\be
\dot M_z =-\bm\nabla\cdot \bm J_s - \frac{\partial R} {\partial \mu}+\bm\nabla\frac{\partial R} {\partial \bm\nabla \mu},
   \ee{canMt}
\be
 \dot\varphi  =- \gamma \mu +\frac{\partial R} {\partial \bm (\bm\nabla\cdot \bm J_s)},
    \ee{canPt}
where the spin chemical potential and the superfluid spin current, 
\be
\mu={\delta F\over \delta M_z},~~\bm J_s = \gamma {\partial F\over  \partial \bm \nabla \varphi},
    \ee{}
are determined by derivatives of the free energy
\be
F={\cal H}+\Omega-TS. 
    \ee{HamAnt}
The spin conservation law forbids the term  ${\partial R/\partial \mu}$  in the continuity equation (\ref{canMt}), because it is not a divergence of some current. Thus, the dissipation function is compatible with the  spin conservation law  if it depends only on the gradient of the spin chemical potential $\mu$, but not on $\mu$ itself. This does not take place in the LLG theory with the Gilbert damping discussed in the Appendix.  The analysis of this section assumed the spin conservation law and corresponded to the dissipation function 
\be 
R =  {\chi D\over 2 } \bm \nabla \mu^2-{ D\over 2 T}{M_\perp^2\over M_z} \bm \nabla \mu\cdot \bm \nabla T
+{  \zeta \over 2\gamma AM_\perp^2}\left(\bm\nabla\cdot \bm J_s\right)^2.
   \ee{Rllgt}
In general the dissipation function contains also the term $\propto \bm \nabla T^2$ responsible for the thermal conductivity. But it is important only for the heat balance equation, which was not considered here. 

If the temperature does not vary in space, then the only temperature effect  is a  correction to the spin chemical potential. This does not affect the basic feature  of superfluid spin transport: there is no gradient of the chemical potential in a stationary current state, and all dissipation processes are not effective except for the relativistically small spin Bloch relaxation.  If there is spatial variation of temperature, then  the spin chemical potential also varies in space. One can find its gradient  by exclusion of $\bm\nabla\cdot \bm J_s$ from Eqs.~(\ref{canMt}) and (\ref{canPt}):
\be
\bm \nabla   \mu={\bm \nabla  M_z\over \chi}= -{ D \zeta \over 2\gamma^2  A M_z  T}\bm \nabla (\bm \nabla ^2 T).
   \ee{grT}
Note that the spin chemical potential gradient is proportional not to the first but to the third spatial derivative of the temperature. The constant temperature gradient does not produce spatial variation of the chemical potential. This is an analog of the absence of thermoelectric effects proportional to the temperature gradients in superconductors.\cite{Ginz} Naturally the effect produced by higher derivatives of the temperature is weaker than produced by the first derivative.  

The nonuniform correction to the spin chemical potential  strongly depends on temperature. Assuming the $T^4$ dependence of the dissipation parameters $D$ and $\zeta$ in \eq{alp} the coefficient before the temperature-gradient term in \eq{grT} is proportional to $T^8$. Now the spin diffusion current $-\chi D \bm \nabla   \mu$ does not disappear in the equation (\ref{canMt}) of continuity for the spin, but it is proportional to $T^{12}$. 

Earlier \citet{Zhang} used the Boltzmann equation for derivation of the spin diffusion coefficient and the Bloch relaxation time in an isotropic ferromagnet in a constant magnetic field. We derived the spin diffusion and the second viscosity coefficients in an easy-plane ferromagnet with  different spin-wave spectrum. Two-fluid effects in easy-plane ferromagnets were investigated by  \citet{TwoFlu}. They solved the Boltzmann equation using the equilibrium magnon distribution function with  nonzero chemical potential of magnon (do not confuse it with the spin  chemical potential introduced in the present paper).  In contrast, we assumed complete thermalization of the magnon distribution when the magnon chemical potential vanishes. The thermalization assumption is questionable in the transient layer near the interface through which spin is injected,  
and in this layer the approach \citet{TwoFlu} may become justified. The transient layer is discussed in the next section. 

\section{Transient (healing) layer near the interface injecting spin} \label{Heal}

Injection of spin from a medium without   spin superfluidity  to a medium with  spin superfluidity may produce not only a superfluid spin current  but also a spin current of incoherent magnons.   But at some distance from the interface between two media, which will be called the conversion healing length, the spin current of incoherent magnons (spin diffusion current) must inevitably transform to superfluid spin current, as we shall show now.

We return back to Eqs.~(\ref{EmB}) and (\ref{EpB}) but now  we  neglect the relativistically small Bloch spin relaxation (the term $\propto 1/T_1$). In Sec.~\ref{SpTr}  we considered  the stationary solution of the these  equations with constant magnetization and absent spin diffusion current. But it is not the only stationary solution. Another solution is an evanescent mode 
$M'_z \propto   \nabla \varphi \propto  e^{-x/\lambda}$, where 
\be
\lambda= \sqrt{  \chi D \zeta \over \gamma {\cal A}  }
  \ee{HL}
is the conversion healing length. We look for superposition of two solutions, which satisfies the condition that the injected current $J_0$  transforms to the spin diffusion current, while the superfluid current vanishes at $x=0$:
\be
J_0 = -D\nabla_x M'_z(0),~~\nabla_x \varphi(0)=0.
    \ee{}
This superposition is 
\be
M'_z(x) =M'_z+{\lambda  J_0\over D}e^{-x/\lambda},~~\nabla_x\varphi (x) ={J_0\over {\cal A}}(1-e^{-x/\lambda} ),
     \ee{interf}   
where $M'_z$ in the right-hand side is a constant magnetization far from the interface $x=0$. Thus, at the length $\lambda$ the spin diffusion current $J_d$ drops from $J_0$ to zero, while the superfluid spin current grows from zero to $J_0$ and remains at larger distances constant. 

As pointed out in the end of Sec.~\ref{SpTr}, the phenomenological equations  (\ref{EmB}) and (\ref{EpB}) were derived assuming that the spin chemical potential $ \mu=M_z'/\chi -H$ does not depend on gradients $\bm \nabla M_z$. However,  the dissipation coefficients $D$ and $\zeta$ decrease very sharply with temperature, and the conversion healing length eventually becomes much smaller than the scale $\xi_0$ [see \eq{xi0}], when the dependence of the free energy and  the spin chemical potential on the gradients $\bm \nabla M_z$ becomes important. But in fact adding $\bm \nabla M_z$-dependent terms into the 
expression for $\mu$,
\be 
 \mu = {M_z\over \chi}- H -{A M^2  \bm \nabla^2 M_z\over  M_\perp^2},
     \ee{muM}
does not affect the expression (\ref{HL}) for the healing length. The generalization of the analysis  reduces to replacing of $M'_z$ in   Eqs.~(\ref{EmB}), (\ref{EpB}),  and (\ref{interf}) by  $\chi \mu$.

Transformation of the injected incoherent magnon spin current to the superfluid spin current is not the only transient process near the interface between media with and without spin superfluidity. Even in the absence of spin  current the interface may affect  the equilibrium magnetic structure. For example,  the interface can induce anisotropy different from easy-plane anisotropy in the bulk. Then the crossover from surface to bulk anisotropy occurs at the healing length of the order of the correlation length $\xi_0$ determined by \eq{xi0} in ferromagnets, or the correlation length $\xi$ determined by \eq{xi} in antiferromagnets.  The  similar healing length was suggested for ferromagnets by \citet{Tserk} and for  antiferromagnets by \citet{Halp} although using different arguments.  

The expression (\ref{HL}) for $\lambda$ was derived within hydrodynamics with dissipation. At distances shorter than the  mean-free path incoherent magnons are in the ballistic regime and cannot converge to the superfluid current, since conversion is impossible without dissipation. Altogether this means that the real healing length at which the bulk superfluid spin current state is formed cannot be less than the longest from three scales: $\lambda$, $\xi_0$, and the magnon mean-free path  $c_s \tau$. Apparently at  low temperatures and weak magnetization $M_z$ the latter is the longest one from three scales. However, close to the phase transition  to the easy-axis anisotropy ($M_z=M$) the coherence length $\xi_0$ diverges and becomes the longest scale. 

Solving the Boltzmann equation we  assumed complete thermalization of the magnon distribution.  At low temperatures when magnon-magnon interaction is weak the length at which thermalization occurs essentially exceeds the mean-free path on defects. It could be that the healing length  would grow up to the thermalization length. This requires a further analysis.

\section{Magnetic vortex in an easy-plane antiferromagnet} \label{AFV}

\begin{figure}[t]
\includegraphics[width=.4\textwidth]{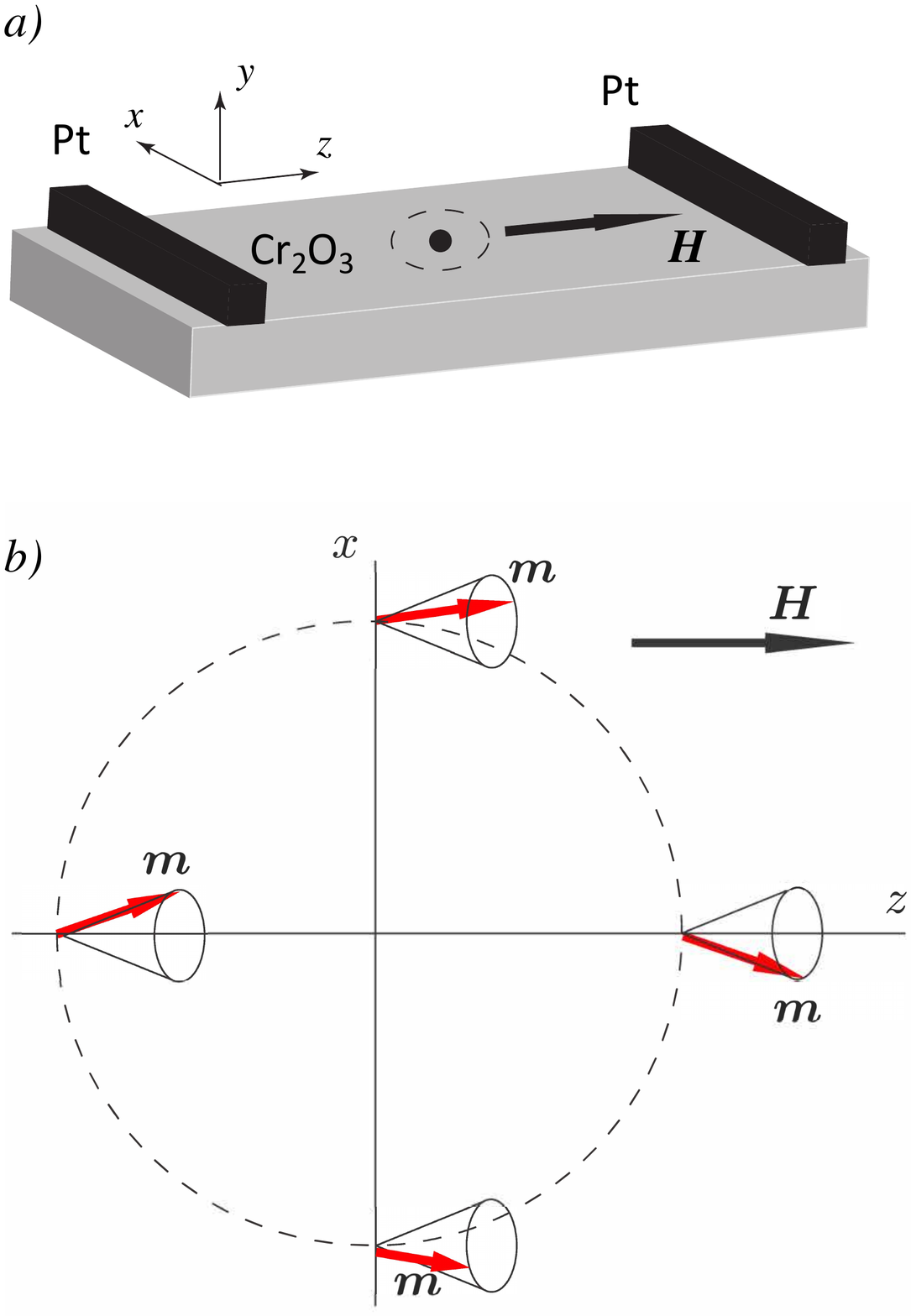}
\caption[]{ Precession of magnetization $\bm m$ around the direction of the magnetic field $\bm H$ along the path around the vortex axis. (a) The geometry of the experiment\cite{WeiH} with the magnetic field (the axis $z$) in the plane of  the Cr$_2$O$_3$ film. The vortex axis is normal to the film (the axis $y$). (b) Precession of the magnetization $\bm m$ is shown in the plane $xz$ (the plane of the film). The path around the vortex axis (dashed lines) is inside the vortex core where the total magnetization is not parallel to $\bm H$ ($\theta\neq 0$).}
\label{f3}
\end{figure}

 Let us consider structure of an axisymmetric vortex in an antiferromagnet with one quantum of circulation of the angle $\varphi_0$ of rotation around the vortex  axis.
Now we consider the geometry of the experiment\cite{WeiH} when the magnetic field $\bm H$ (the axis $z$) is in  the film plane. The vortex axis is the axis $y$ normal to the film plane (Fig,~\ref{f3}a). The azimuthal component of the angle $\varphi_0$  gradient is
\be
\nabla \varphi_0 ={1\over r}.
      \ee{}
At the same time $\varphi =0$ and $\theta_0$ is small. Then the Hamiltonian (\ref{H}) transforms to
\be
{\cal H}={2M^2\over \chi}\theta_0^2 -2H M\cos \theta  \theta_0 
+A_-M^2\left( {\cos ^2\theta\over r^2}  +\nabla \theta^2\right).
   \ee{}
Minimization with respect to small $\theta_0$ yields
\be
\theta_0={\chi H\cos \theta\over 2M},
   \ee{}
and finally the Hamiltonian is
\be
{\cal H}= {-\chi H^2\cos^2 \theta \over 2}
+A_-M^2\left( {\cos ^2\theta\over r^2}  +\nabla \theta^2\right).
   \ee{}
The Euler--Lagrange equation for this Hamiltonian describes the vortex structure in polar coordinates: 
\be
{d^2 \theta\over dr^2}+{1\over r}{d \theta\over dr}-{\sin 2\theta\over 2}\left({ 1 \over \xi^2} -{1 \over r^2}\right) =0,
   \ee{}
where the correlation length $\xi$ is given by \eq{xi} and determines the size of the vortex core.  

The vortex core has a structure of a skyrmion, in which the total weak magnetization deviates from the direction of the magnetic field $\bm H$ ($\theta\neq 0$). The component of magnetization transverse to the magnetic field is 
\be
m_\perp=  {\gamma H \sin 2\theta \over 2}.
   \ee{} 
The transverse magnetization creates stray magnetic fields at the exit of the vortex line from the sample.  Figure~\ref{f3} shows variation of the magnetization inside the core along the path around the vortex axis parallel to the axis $y$. Along the path the  magnetization $\bm m$ revolves around the  direction of the magnetic field forming a cone.  The precession in space creates an oscillating $y$ component of magnetization  $m_y=m_\perp(r) \sin \phi$, where $\phi $ is the azimuthal angle at the circular path around the vortex line. This produces surface magnetic charges $4\pi m_y$ at the  exit of the vortex to the boundary separating the sample from the vacuum. These charges generate  the curl-free stray field $\bm h=\bm \nabla \psi$.  At distances from the vortex exit point much larger that the core radius the stray field is a dipole field with  the scalar potential 
\bem
 \psi(\bm R)  ={\pi\chi H\over 2} { (\bm  R\cdot\bm n)\over R^3} \int _0^\infty \sin 2\theta(r) r^2\,dr
 \nonumber \\
 =1.2\pi\chi H\xi^3{ (\bm  R\cdot\bm n)\over R^3}={1.2\pi \chi c_s^3\over \gamma^3 H^2}{ (\bm  R\cdot\bm n)\over R^3}.
     \eem{}
Here $\bm R(x,y,z) $ is the position vector with the origin in the  vortex exit point and $\bm n$ is a unit vector in the plane $xz$ along which the surface charge is maximal ($\phi=\pi/2$).  In our model the direction of $\bm n$ is arbitrary, but it will be  fixed by spin-orbit interaction or crystal magnetic anisotropy violating invariance with respect to rotations around the axis $z$. These interactions were ignored in our model.  In principle, the stray field can be used for detection of vortices nucleated at spin currents approaching the critical value.

\section{Discussion and summary} \label{DS}

The paper analyzes the  long-distance superfluid spin transport.  The superfluid spin transport does not require a gradient of the spin chemical potential (as the electron supercurrent in superconductors does not require a gradient of the electrochemical potential). As result of it,  mechanisms of dissipation are suppressed except for weak Bloch spin relaxation. Other dissipation mechanisms affect the spin transport only at  the transient (healing) layer close to the interface through which spin is injected, or in nonstationary processes.

The paper calculates  the Landau critical spin phase gradient in a two-sublattice antiferromagnet when the easy-plane topology of the magnetic  order parameter is provided not by crystal magnetic anisotropy but by an external magnetic field. This was the case realized in the experiment by \citet{WeiH}.  For this goal it was necessary to derive the spectrum of collective modes (spin waves)  in spin current states.
The Landau instability destroying spin superfluidity sets on not in the Goldstone gapless mode as in easy-plane ferromagnets but in the gapped mode, despite that at small spin currents the latter has energy larger than the Goldstone mode. 

The paper analyzes  dissipation processes determining  dissipation parameters  (spin diffusion and second viscosity coefficients)   by solving the Boltzmann equation for magnons scattered by defects. The two-fluid theory similar to the superfluid two-fluid  hydrodynamics  was suggested.   It is argued that the LLG theory with the Gilbert damping parameter is not able to properly describe dissipation in easy-plane magnetic insulators. Describing the whole dissipation by a single Gilbert parameter one cannot differentiate between  strong processes connected with high exchange energy (e.g., spin diffusion) and weak processes connected with spin-orbit interaction (Bloch spin relaxation), which violate the spin conservation law.     

The formation of the superfluid spin current in the transient (healing) layer near the interface through which spin is injected was investigated. The width of this layer (healing length) is determined by processes of dissipation, and at low temperatures can reach the scale of relevant mean-free paths of magnons including those at which the magnon distribution is thermalized.

The structure of the magnetic vortex in the geometry of the experiment on Cr$_2$O$_3$ is investigated. In the vortex core there is a magnetization along the vortex line, which is normal to the magnetic field. This magnetization produces magnetic charges at the exit of the vortex line from the sample. The magnetic charges create a stray dipole magnetic field, which probably can be used for detection of vortices.

Within the developed two-fluid theory the paper addresses the role of the temperature variation in space on the superfluid spin transport. This is important because in the experiment of \citet{WeiH} the spin is created  in the Pt injector by heating (the Seebeck effect). Thus the spin current to the detector is inevitably accompanied by heat flow. The temperature variation produces the bulk Seebeck effect, which  is estimated to be rather weak at low temperatures. However, it was argued\cite{Lebrun} that probably \citet{WeiH} detected a signal not from spin coming from the injector but from spin produced by the Seebeck effect at the interface between the heated antiferromagnet and the Pt detector. Such effect has already been observed for antiferromagnet Cr$_2$O$_3$.\cite{Seki} If true, then \citet{WeiH} observed not long-distance spin transport but long-distance heat transport. It is not supported by the fact that  Yuan {\em et al.} observed a threshold for  superfluid spin transport at low intensity of injection, when according to the theory\cite{Adv} violation of the approximate spin conservation law becomes essential. Investigation of superfluid  spin transport at low-intensity injection is more difficult both for theory and experiment. But the existence of the threshold is supported by extrapolation of the detected signals from high-intensity to  low-intensity injection. According to the experiment, the signal at the detector is not simply proportional  to the squared electric current $j^2$ responsible for the Joule heating in the injector, but to  $j^2+ a$. The offset $a$ is evidence of the threshold, in the analogy with the offset of $IV$ curves in the mixed state of type II superconductors determining the critical current for vortex deepening. With all that said, the heat-transport interpretation cannot be ruled out and deserves further investigation. According to this interpretation, one can see the signal observed by \citet{WeiH} at the detector even  if the Pt injector is replaced by a heater, which produces the same heat but no spin. An experimental check of this prediction would confirm or reject the heat-transport interpretation.   

Let us make some numerical estimations for Cr$_2$O$_3$ using the formulas of the present paper.  It follows from neutron scattering data\cite{Samu} that the spin-wave velocity is $c_s=8\times 10^5$ cm/sec. According to \citet{Foner}, the magnetization of sublattices is $M=590$ G and the magnetic susceptibility is $\chi= 1.2\times 10^{-4}$. Then the total magnetization $m_z=\chi H$ in the magnetic field $H=9$ T used in the experiment is about 10 G, and 
the canting angle $\theta_0 =m_z /2M  \approx 0.01$ is small as was assumed in our analysis. The correlation length (\ref{xi}), which  determines vortex core radius,  is about $\xi \approx 0.5\times 10^{-6}$ cm. The stray magnetic field produced by magnetic charges at the exit of the vortex line from the sample is $10 (\xi^3/R^3)$ G, where $R$ is the distance from the vortex exit point. The task to detect such fields does not look easy, but it is hopefully possible with modern experimental techniques.

\begin{acknowledgments}
I thank   Eugene Golovenchits, Wei Han, Mathias Kl\"{a}ui, Romain Lebrun,  Allures Qaiumzadeh, Victoria Sanina, So Takei, and Yaroslav Tserkovnyak for fruitful discussions and comments. 
   \end{acknowledgments}

\appendix*
 
\section{Dissipation in the LLG theory} 

For  ferromagnets  the LLG equation taking into account dissipation  is
\be
{d\bm M\over dt}=\gamma \left[ \bm H_{eff}  \times \bm M\right]+ {\alpha\over M}\left[\bm M \times {d\bm M\over dt}\right],
       \ee{LLGd}
where  $\alpha$ is the dimensionless Gilbert damping parameter. For small  $\alpha$ this equation is identical to the equation with the  Landau--Lifshitz damping term:
\be
{1\over \gamma}{d\bm M\over dt}=\left[\bm M  \times {\delta {\cal H}\over \delta \bm M}\right]+ {\alpha\over M}\left[\bm M \times  \left[\bm M  \times {\delta {\cal H}\over \delta \bm M}\right]\right].
       \ee{LLd}
Transforming the vector LLG equation to the equations for two Hamiltonian conjugate variables, the  $z$ component $M_z$ of magnetization and the angle $\varphi$ of rotation around the $z$ axis, one obtains Eqs.~(\ref{canMt}) and (\ref{canPt}) without the term $\bm\nabla(\partial R/\partial \bm\nabla \mu$) and with the dissipation function  
\be 
R =  {\alpha \gamma  M_\perp^2\over 2 M}  \mu^2+{\alpha M \over 2 M_\perp^2}\left(\bm \nabla \cdot \bm J_s \right)^2,
   \ee{Rllg}
which depends on the spin chemical potential $\mu$ itself, but not on its gradient. Meanwhile, according to the two-fluid theory of Sec.~\ref{Boltz}, the $\bm\nabla \mu$-dependent term in the dissipation function was responsible for the spin-diffusion term in the continuity equation for $M_z$. Indeed, at derivation of the continuity equation (\ref{EmB})  from the LLG theory under the assumption  that $\mu \approx M'_z/\chi =M_z/\chi-H$ the spin diffusion term $\propto D$ does not appear. The term does appear only if $\mu$  in the dissipation function  (\ref{Rllg}) is determined by the more general expression (\ref{muM}) taking into account the dependence on $\bm \nabla M_z$.
 Then one obtains Eqs.~(\ref{EmB}) and (\ref{EpB})   with  the  equal spin diffusion and spin second viscosity coefficients 
\be
D=\zeta= \alpha \gamma  MA,
      \ee{Dzeta}
and the inverse Bloch relaxation time
\be
{1\over T_1} ={\alpha \gamma M_\perp^2\over \chi M}.
     \ee{} 
The outcome looks bizarre. The spin diffusion emerges from the $\mu$-dependent term in the dissipation function, which is incompatible with the spin conservation law, as if the spin diffusion is forbidden by  the spin conservation law. Evidently this conclusion is physically incorrect.  
Moreover, in the analogy of  magnetodynamics  and superfluid hydrodynamics the magnetization $M_z$ corresponds to the fluid density. In hydrodynamics the fluid density gradients are usually not taken into account in the Hamiltonian and in the chemical potential since they become important only at small scales beyond the hydrodynamical approach. This does not rule out the diffusion process. Similarly, one should expect that it is possible to ignore the magnetization gradients in the spin chemical potential either. It is strange that the spin diffusion becomes impossible in the hydrodynamical limit.

According to the Noether theorem the total magnetization along the axis $z$   is conserved if the Hamiltonian is invariant with respect to rotations around the axis $z$ in the spin space.  The Landau--Lifshitz theory of magnetism\cite{LLelDy} is based on the idea that the spin-orbit interaction, which breaks rotational symmetry in the spin space and therefore  violates the spin conservation law, is relativistically small compared to the exchange interaction because the former is inversely proportional to the speed of light. So, although the spin conservation law is not exact,  it is a good approximation (see Sec.~\ref{Intr}). Then  the spin Bloch relaxation term $\propto 1/T_1$, which violates the spin conservation law, must be proportional to a small parameter inversely proportional to the speed of light and cannot be determined by the same Gilbert parameter as other dissipation terms, which do not violate the spin conservation law

The insufficiency of the LLG theory for description of dissipation was discussed before, but mostly at higher temperatures. It was suggested to replace of the LLG equation by the Landau--Lifshitz--Bloch equation, in which the Bloch longitudinal spin relaxation is present explicitly (see, e.g., Ref.~\onlinecite{LLB} and references to earlier works therein).  Our analysis shows that the problem exists also at low temperatures.

%





\end{document}